\documentclass[utf8]{FrontiersinHarvard} 

\usepackage{url,hyperref,lineno,microtype,subcaption}
\usepackage[onehalfspacing]{setspace}
\usepackage{aas_macros}
\usepackage{xcolor}
\usepackage{ulem}
\usepackage{newtx}

\def\keyFont{\fontsize{8}{11}\helveticabold }
\def\firstAuthorLast{Blaschke {et~al.}} 
\def\Authors{David Blaschke\,$^{1,2,3,*}$, Friedrich K. R\"opke\,$^{4,5,6}$ and Gerd R\"opke\,$^{7}$}
\def\Address{
$^{1}$Helmholtz-Zentrum Dresden-Rossendorf (HZDR), Bautzener Landstrasse 400, 01328 Dresden, Germany \\
$^{2}$Center for Advanced Systems Understanding (CASUS), Untermarkt 20, 02826 G\"orlitz, Germany \\
$^{3}$Institute of Theoretical Physics, University of Wroclaw, Max Born place 9, 50-204 Wroclaw, Poland \\
$^{4}$Astronomisches Recheninstitut, Zentrum f\"ur Astronomie der Universit\"at Heidelberg, M{\"o}nchhofstr.\ 12--14, 69120 Heidelberg, Germany\\
$^{5}$Institut f\"ur Theoretische Astrophysik, Zentrum f\"ur Astronomie der Universit\"at Heidelberg, Philosophenweg 12, 69120 Heidelberg, Germany\\
$^{6}$Heidelberger Institut f\"ur Theoretische Studien, Schloss-Wolfsbrunnenweg 35, 69118 Heidelberg, Germany\\
$^{7}$Institute of Physics, University of Rostock, Einstein-Str. 23, 18057 Rostock, Germany}

\def\corrAuthor{David Blaschke}

\def\corrEmail{david.blaschke@uwr.edu.pl}

\definecolor{frcolor}{rgb}{0,0.5,0}

\definecolor{dbcolor}{rgb}{0,0,0.5}

\begin{document}
\onecolumn
\firstpage{1}

\title[Universality of heavy r-process elements]
{Universality and variability of the heavy r-process element abundance pattern from a nonequilibrium approach}

\author[\firstAuthorLast ]{\Authors} 
\address{} 
\correspondance{} 

\extraAuth{}

\maketitle
\date{\today}

\begin{abstract}

\section{}
A striking feature in the observed chemical composition of the majority of stars is the universality of the relative abundances of the heavy elements, although some outliers exist.
We demonstrate that a nonequilibrium freeze-out approach provides a natural way of accounting for the typical abundance pattern and its variation. Here, we use a phenomenological method to characterize the coarse-grained distribution of heavy $r$-process elements in several astrophysical objects. The Lagrange parameters show only minor fluctuations when comparing different stars. 
Larger deviations are observed in stars with low metallicity.
The variations in the Lagrange parameters for these stars are presented.
The determination of the Lagrange parameters can be instrumental in identifying possible sources for the formation of heavy elements.
In particular, density fluctuations are considered as a source for the production of heavy elements in the early Universe.

\tiny
 \keyFont{ \section{Keywords:} nucleosynthesis, $r$-process, solar abundances, mass fractions of heavy nuclei, galactic chemical evolution} 
\end{abstract}

\section{Introduction}
\label{sec:intro}

Over the past decades, spectroscopic observations have provided detailed abundances of chemical elements in various stars.
For instance, the \textit{Satellites Around Galactic Analogs} (SAGA) survey \citep[see][and references therein]{2024ApJ...976..118G}
significantly expands on the rich observational datasets from satellite systems in the Milky Way. 
A long-standing questions is how and where the observed heavy elements with charge number $Z\,{>}\,28$ are formed. 
For elements below the Fe--Ni range, exothermic  burning processes can occur. For the heavy elements, fusion processes are energetically suppressed. 
Aside from decay [and the so-called $s$-process], reaction kinetics for forming of these heavy elements has been associated with rapid neutron-captures -- the so-called $r$-process -- for which a special environment is required \citep{Burbidge:1957vc}.
{For a detailed discussion of the $r$-process, we refer to the reviews of \citet{1991PhR...208..267C,2007PhR...450...97A}, and \citet{2011PrPNP..66..346T}.}
The  production sites of the observed distributions of heavy elements are still heavily debated \citep[for a recent review, see][]{2021RvMP...93a5002C}.

Here, we do not intend to solve the problem of the origin of the heavy elements, but study the question whether the observational information about the distribution of the heavy elements can be cast in a minimal set of three Lagrange parameters (related to the temperature and the chemical potentials of neutrons and protons) which characterize a freeze-out scenario  for heavy-element formation \citep{Roepke:2024fzj}. 
This is a phenomenological approach to characterize general properties of the observed distributions. 
In general, the temporal evolution of the elemental abundances is described by reaction kinetics. 
However, to solve these differential equations, the initial state is required,
and the thermodynamic state of the site must be specified. 
For this, the determination of the Lagrange parameters is of use. 
The aim of our work is that the universal pattern observed in many stars provides an approximation to an initial condition. 
The initial state can be characterized by a quasi-equilibrium state with few Lagrange parameters. 
This initial state is denoted as freeze-out state. 
The subsequent evolution cannot be described by a quasi-equilibrium state.
The distribution can later be changed, and this temporal evolution is then described by reaction kinetics.

For the solar system, the abundance of isotopes $\{A,Z\}$, i.e.\ nuclei with mass number $A$ and charge number $Z$,
is well known, see \citet{2021SSRv..217...44L}.
For other stellar objects, the chemical composition (mostly elemental and not isotopic abundances) has been deduced from the spectral analysis of the emitted light,
i.e. the elemental abundances $Y_Z =\sum_A n_{A,Z}/n_B$, where $n_{A,Z}$ denotes the density of nuclei $\{A,Z\}$, and $n_B$ the baryon number density.
An interesting observation when comparing the composition of the sun to that of other stellar objects is that the relative abundances of $r$-process elements show a very similar pattern. 
This is shown, for instance in Figure 3 of \citealt{2021RvMP...93a5002C}.
An interpretation of this finding is a ``robust'' $r$-process that pollutes stellar material early in the evolution of our Galaxy -- perhaps taking place in a single astrophysical site -- as opposed to a random superposition of yields from various sources \citep{1999ApJ...521..194C}. 
The phenomenon is usually referred to as the so-called \textit{universality} of the $r$-process \citep[see also][]{2022ApJ...936...84R}.

To compare the relative proportion of chemical elements $i,j$ for different objects $A,B$, we consider the logarithm $\log[Y_{Z_i}^AY_{Z_j}^B/(Y_{Z_j}^AY_{Z_i}^B)]$ which takes the value zero if the relative proportions in object $A$ and $B$ are identical. 
Here, $Z_i$ denotes a chemical element, and we consider only heavy elements with charge number $Z \ge Z_{\rm heavy}\approx 30$.
Often, the solar abundance is taken as reference, and the relative ratios $[Z_i/Z_j]=\log [Y_{Z_i}Y_{Z_j}^\odot/(Y_{Z_j}Y_{Z_i}^\odot)$ are introduced, that can be considered as a measure of the universality. If in both objects $A,B$ the ratio of the heavy element abundances are identical, we have $[Z_i/Z_j]^A=[Z_i/Z_j]^B= {\rm const}$, independent of $i,j$.

The pattern of the heavy element distribution seems to be uniform for the majority stars in the nearby parts of the Milky Way;
the same pattern is also observed in metal-poor stars  in the Galactic halo, see \citet{2021RvMP...93a5002C}. These metal-poor stars are considered to be very old, representing the composition in the early stage of the Universe (hence the associated term ``{\it galactic archeology}''). 
However, exceptions to the uniformity of the observed abundance patterns are known, which will be discussed below.
 
In the present work, we aim to  determine values of Lagrange parameters that characterize these deviations from the solar distribution. 
Such deviations encompass abundances that are overall scaled up or down but still follow the same pattern -- which would still be consistent with universality -- but also changes in the pattern itself -- indicating a loss of universality.

After {summarizing the observational results on the $r$-process abundance pattern in stars (Section~\ref{sec:universality_obs}) and} considering the distribution function in Section~\ref{distributions} (improved $T$-dependence of the shell correction contribution), we discuss universality for the Sun and other stars in Section \ref{universality}. 
The relevance of metal-poor stars for the early Universe is discussed in Section \ref{sec:3}, and finally, we draw conclusions in Section \ref{conclusion}.
The problem of where such Lagrange parameters occur is left to future work.

\section{Universality of the $r$-process abundance pattern and deviations from it}
\label{sec:universality_obs}

In this section, we summarize recent observational findings concerning the abundance distribution of heavy elements in stars. In particular, we focus on the question of how strong the indication of universality is. 

Most low-metallicity stars are observed to have a similar abundance pattern for $r$-process elements \citep{2023MNRAS.524.5607S,2024arXiv241115415F, 2025A&ARv..33....2B}. This is also supported by \citet{2018ARNPS..68..237F}, who points out that Pb, Th and U still provide the tightest observational constraints on the poorly understood actinide production.
Comprehensive abundance data on dwarf and giant stars in the Galactic halo (thick and thin disks) were published by \citet{2012A&A...545A..31H}. Their findings (large star-to star scatter) seem to favor an {\it early inhomogeneous interstellar medium}. 
We discuss these abundances in Sect. \ref{universality}.
\citet{2018ApJ...865..129R,2022ApJS..260...27R} present the so-far most complete chemical inventory for the metal-poor star HD~222925 and note that the $r$-process elements in the range $Z \ge 56$ (including the third $r$-process peak) show a near-perfect match to the solar pattern when scaled to the Eu abundance.
The distribution of the heavy elements shows a behavior similar to the giants reported by \citet{2012A&A...545A..31H}, see Sect. \ref{universality}.

Deviations from the pattern observed in the majority of stars are a strong reduction of the heaviest elements in the stars analyzed by \citet{2007ApJ...666.1189H}, see also Fig.~4 of \citet{2021RvMP...93a5002C}. 
We discuss these drop-off distributions in Sect. \ref{universality}.
In contrast, there is an overabundance of the heaviest elements, such as Th and U, as compared to the other $r$-process elements in several stars with metallicities ${\rm[Fe/H]}\, {\approx}\, -3$ \citep{2021RvMP...93a5002C}, usually referred to as an ``{\it actinide boost}''. \citet{2021RvMP...93a5002C} attribute this finding to either an $r$-process contributing to very early galactic evolution 
or {\it varying conditions} in the sites of the $r$-process.

A recent overview of observations regarding the third $r$-process peak was published by \citet{2025A&A...693A.294A}.
Comparing several stars, they show that Pt, but also Ir and Os,  are overproduced in contrast to 
Hf, in particular for stars with low Eu abundances. 
They point out that none of the current models can explain the observed abundances in the third peak and speculate about an additional $r$-process active at low metallicities (i.e.\ very early epochs) that favors production of elements in the third peak while contributing little to Eu.
The {\it non-robustness} of the $r$-process for the actinide-boosted stars was also discussed in \citet{Eichler:2014kma} and \citet{2019ApJ...881....5H}. 

As discussed in Sect. \ref{distributions}, $\alpha$-decay and fission processes must be considered which give an additional population of the heaviest elements.
It was found that heavy element distributions show signatures of fission \citep{2011A&A...534A..60B}, see also \citet{ernandez2023a} and  \citet{2023Sci...382.1177R}.

After the discussion of universality and deviations in both directions, under- and over-population of the heaviest elements,
we would like to mention another interesting point related to the abundance of the heaviest elements --  {\it cosmochronology}.
The search for low-metallicity stars should answer the question of the early appearance of heavy elements in the Universe,  see \citet{2025A&ARv..33....2B}.
For instance, abundance ratios of actinides can be used a chronometer pairs to calculate the age of stars,
see \citet{2025ApJ...984L..43L}.
The inferred age of the observed star J0804+5740 is consistent with the age of the Universe indicating heavy elements to be present already in the early Universe, see also
\citet{2024ApJ...971..158R,2025A&A...697A.127H,2025ApJ...988...22H}. 
We discuss this topic in Sect. \ref{sec:3}.

We collected some examples of observations of the chemical composition of stars to discuss the universality of heavy element distribution.
Universality is often confirmed in good approximation, but there exist also deviations which should be understood.
However, as pointed out in the references given above, there are many open questions about the origin and the site where the heavy elements are formed.

Current approaches that model the origin of heavy elements  in the galactic chemical evolution start from a state without heavy elements.
Assuming the homogeneous big bang scenario \citep{Burbidge:1957vc}, the primordial composition at nucleosynthesis ($t_{\rm nucl} \approx 100$ s after big bang) has only H, He, and a small amount of Li. 
Stars, made of this primordial matter, are denoted as population III. 
All heavy elements are formed later on, in particular the so-called heavy $r$-process elements. 
Different sites have been considered, where such rapid neutron capture processes can occur, for instance supernova explosions and mergers of compact objects (neutron stars, black holes) .

Models have been worked out to simulate the heavy-element production by these processes, see \citet{2021RvMP...93a5002C} for a review. 
Extended hydrodynamical simulations have been performed, and postprocessing the formation of the heavy elements is described by nuclear reaction network simulations. 
The output of the heavy element production is determined by the particular astrophysical conditions such as the neutron star masses in the merging binary system and the trajectory of a mass element in the phase space.
This way, it is not always possible to produce also the elements in the lead region (the third peak around $A \approx 195$) or the actinides so that the universality remains a puzzle.

A solution was proposed in \citet{deJesusMendoza-Temis:2014owk} where the nuclear robustness of the $r$ process in neutron-star mergers was explained by repeated fission cycles, and the final abundance distribution is not strongly dependent on the initial astrophysical conditions.
With special assumptions, they showed that the observed pattern of 
heavy element distribution can be reproduced by simulations.
However, they concluded that the actual astrophysical site of the $r$ process is not yet known.

A superposition of different sources to explain universality and its deviations was proposed by \citet{2015MNRAS.448..541J}.
Recently, a complete survey of $r$-process conditions and the un-robustness of $r$-process has been published by \citet{2025arXiv250600092K}.
They found that expensive hydrodynamic simulations of extreme environments such as neutron star mergers show that a wide range of conditions produce very similar abundance patterns explaining the observed robustness of the $r$ process between the second and third peak. However, it was not possible to find a single condition that produces the full $r$ process from the first to the third peak. 
Instead, a superposition of different conditions or components is required to reproduce the typical $r$ process pattern as observed in the solar system and very old stars. In their work, it was not aimed to link the different conditions to a given astrophysical site.

In this work, we analyze the chemical composition of various objects and ask whether we can characterize the distribution of heavy elements using the heavy-element freeze-out (HEFO) Lagrange parameters $\lambda_T, \lambda_n, \lambda_p$ \citep{2025Univ...11..323R,Roepke:2024fzj},  
which are the nonequilibrium generalizations of the equilibrium parameters $T,\mu_n,\mu_p$. 
Our approach is phenomenological, we extract the properties of the source at freeze-out from the data, but we give not a microscopic approach which describes the dynamical process how these freeze-out states can occur.
Our point is to express these "{\it varying conditions}" \citep{2021RvMP...93a5002C} which are introduced to describe individual chemical composition of stars by varying parameter values of Lagrange parameters $\lambda_T, \lambda_n, \lambda_p$.

\section{The heavy element freeze-out approach}
\label{distributions}

The chemical composition of stellar matter is expressed by the mass fraction $X_{AZ}=A n_{AZ}/n_B$ with density $n_{AZ}$ of isotopes $\{A,Z\}$; 
$n_B$ is the baryon number density, and $\sum_{AZ} X_{AZ}=1$.
Detailed results for the solar distribution of isotopes  $X^\odot_{AZ}$ are well known. A related quantity is the abundance of elements, $Y_Z=\sum_A n_{AZ}/n_B$, often used in comparison to the solar ones as $[Z,Z']=\log[Y_Z Y^\odot_{Z'}/Y_{Z'} Y^\odot_{Z}]$.

To describe the temporary evolution of the distribution function, we require a non-equilibrium approach.
For hot and dense matter, a hydrodynamical description is possible, where local thermodynamic equilibrium is assumed.
Correlations and the formation of bound states can occur, but relax quickly to the equilibrium with the local thermodynamic parameters. If the hot and dense matter expands and cools down, the relaxation time for equilibrium can become larger than the rate of the change of the thermodynamic parameters so that the local thermodynamic equilibrium is no longer established, this thermodynamic state freezes out. 
In particular, the equilibrium distribution function of the isotopes freezes out. Of course, there are changes possible also after freeze-out, but these must be described by reaction kinetics.

In the simulation of supernova or merger processes, for expanding hot and dense matter 
hydrodynamic equations based on local thermodynamic equilibrium are used. 
The isotopic distribution is obtained from postprocessing where below a typical temperature (about 0.5 MeV) nuclear reaction networks are used to simulate the evolution of the distribution function.
As starting point for the isotopic distribution function, the nuclear statistical equilibrium (NSE) is used. 

A systematic approach should use a nonequilibrium approach \citep{Roepke:2024fzj}. 
In general, in-medium corrections can be taken into account, with respect to the binding energies of the isotopes as well as to the reaction rates. 
A consistent description of the freeze-out concept can be given using the method of the nonequilibrium statistical operator, see \citep{Roepke:2024fzj}. 
Lagrange parameters $\lambda_i$ are required to determine the non-equilibrium state of the system.
After freeze-out, reaction kinetics determines the evolution of the system, for instance decay processes of excited states of nuclei.
This approach {provides a complete spectrum of element abundances including heavy $r$-process elements as a seed distribution for subsequent postprocessing in stellar and explosive nucleosynthesis.} 
Instead of NSE, slow variables 
are taken into account to characterize the state of the system, and in-medium corrections are considered.

To analyse the heavy element distribution, our assumption is, to start from a hot and dense state of matter (for instance, supernova explosions, neutron star mergers, or other states in the early Universe) and to follow the expansion of hot and dense matter, i.e. the decompression and cooling process. Bound states (nuclei) are formed if the density is smaller than the Mott density. Reactions occur, and detailed balance move the system towards thermodynamic equilibrium. However, if some reactions become slow, the corresponding degrees of freedom freeze out, and the corresponding averages characterize the further evolution like quasi-constants of motion.
Of course, there are also changes of the composition after freeze-out, but these must be described as kinetic processes. The distribution function of the elements is no longer given by the nuclear statistical equilibrium; their temporal evolution is described by reaction kinetics.
Different processes can be considered which change the composition of matter. The decay of excited states ($\gamma$ decay) and the $\beta$ and $\beta^+$ reactions do not change the mass number $A$ of the nuclei. To get rid of these processes, we consider the mass number distribution $X_A=\sum_Z X_{AZ}$. This mass number distribution is changed by emission/absorption of neutrons/protons, $\alpha$ particles, and generally by fusion and fission processes.

To describe the nonequilibrium evolution of the chemical composition, we have to consider different reactions which are relevant for the chemical evolution. We assume that the details of the distribution of elements are formed in a late stage whereas some general features are formed already very early during the chemical evolution process. Our aim is to identify slow variables which can be used to construct the relevant statistical operator. Finer details such as the staggering with respect to $N$ or $Z$ are subject of the very late stage, to be described by detailed nuclear reaction networks.

\subsection{The coarse-grained distribution function}
To identify these gross structures of the distribution function, we consider a coarse-grained distribution \citep{1987PhLB..185..281R}, 
the accumulated mass fraction 
\begin{equation}\label{XA}
\hat X_{\hat A}=\frac{1}{n_B} \sum_{A'=\hat A}^{\hat A+3} A' \sum_{Z,\nu} n_{A',Z,\nu}
\end{equation}
with $\nu$ denoting the excitation state of the isotope $\{A,Z\}$, 
 and the $\hat A$-metallicity
\begin{equation}\label{MA}
M_{\hat A}=\sum _{\hat A' \ge \hat A} \hat X_{\hat A'}~.
\end{equation}
Here, $n_\mathrm{B}$ denotes the baryon number density, and $n_{A,Z,\nu}$ the number density of clusters with mass number $A$ and charge number $Z$. The intrinsic quantum number $\nu$ gives the excitation state of the nucleus $\{A,Z\}$ and $\hat A$ characterizes the group of clusters; it can take values in $[0, 4, 8, 12, \ldots ]$.

We are not dealing with the high abundance of the well-bound $\alpha$ nuclei ($^{12}$C, $^{16}$O, etc.) and the even-odd staggering which are designed only in the late stage of evolution, but merely in the global structure of the distribution function.

In particular, we focus on the heavy elements $A\ge 76$ which are beyond the iron peak. 
While the light elements up to the iron/nickel region are produced steadily in stellar burning processes, the heavy elements are mainly frozen out. 
Special conditions are necessary to run the $r$ or $s$ process which recently are possible, e.g., in SN explosions or NS mergers. 
Heavy elements are observed in various astrophysical objects, but the site where they are formed has not been fully resolved yet, as outlined in the Introduction and in Sect. \ref{sec:universality_obs}.
For a review on the deciphering the origins of the elements through galactic archeology see \cite{2025arXiv250318233F} where further references can be found. 

We have recently published an 
article \citep{Gonin:2025uvc} that was based on the 
concept of HEFO \citep{1987PhLB..185..281R,Roepke:2024fzj}.
As a prerequisite to determine the conditions at which the heavy element distribution is formed, we determine freeze-out conditions which are fitted to the observed heavy element distribution. For the solar distribution, we found 
the Lagrange  parameter values $\lambda_T=5.266\,\mathrm{MeV}$, $\lambda_n=940.317\,\mathrm{MeV}$ and  $\lambda_p=845.069\,\mathrm{MeV}$, 
{which represent the non-equilibrium generalizations of the temperature $T$ and the chemical potentials $\mu_n$ and $\mu_p$ for neutrons and protons, respectively.
HEFO parameter values of temperature $T\approx 5$ MeV, corresponding to $5.8 \times 10^{10}$ K, baryon number density $n_B \approx 0.013$ fm$^{-3}$, corresponding to a mass density of $\varrho=2.2 \times 10^ {13}$ g/cm$^3$, and proton fraction $Y_p=0.13$ are found in simulations of supernova explosions, see \citet{2014EPJA...50...46F,2017PASA...34...67F}, and in the crust of proton-neutron stars, see \citet{2023A&A...672A.160D}.
This HEFO  distribution includes also a large amount of superheavy elements which decay afterwards} by $\alpha$-particle emission or fission. 
Since knowledge about the branching rates of different decay processes of these superheavies is scare, we can only perform crude estimates. However, the value $M_{76}$, the total mass fraction of heavy nuclei, is nearly unchanged with respect to fission and $\alpha$-decay processes so that this value can be considered as a nearly conserved quantity to construct the non-equilibrium statistical operator.

Such quasi-conserved quantities are of interest when describing the nonequilibrium distribution. 
This means that the assumption, often made in simulations using nuclear reaction networks, that above a temperature of about 0.5 MeV local nuclear statistical equilibrium can be assumed, is questionable. At heavy-element freeze-out (HEFO), where the neutron density is going down, the relaxation time for the heavy element metallicity $M_{76}$ becomes too long to arrive at equilibrium.

\begin{figure}
     \begin{center}
 \includegraphics[width=0.6\linewidth]{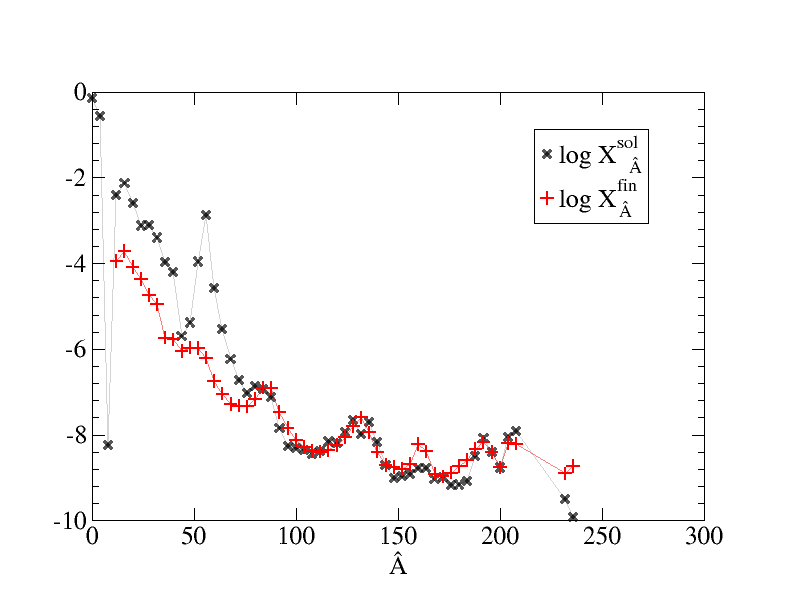}
    \end{center}
     \caption{Fig. 10 from \citet{Gonin:2025uvc}. Accumulated mass fraction $\hat X_{\hat A}$ (red ``$+$''
symbols) for the parameter values $T=5.266\,\mathrm{MeV}$,
$\mu_n=940.317\,\mathrm{MeV}$ and  $\mu_p=845.069\,\mathrm{MeV}$ after
evaporation of neutrons $\hat X'_{\hat A}$ compared with the solar
accumulated mass fraction $\hat X^\odot_{\hat A}$ (black ``$\times$''
symbols. 
In addition to neutron evaporation, leading to  $\hat X'_{\hat A}$ (see \citet{Roepke:2024fzj} for details) nuclei
with $A>212$ are subject to $\alpha$-decay (feeding the region near $A\sim 200$) and fission (feeding the region near $A\sim 160$). 
}
     \label{fig:XA1}
\end{figure}

As shown in Fig. \ref{fig:XA1}, at HEFO the solar heavy element distribution is well reproduced,
while light elements are underproduced. In particular, the iron peak is underproduced by a factor of about $10^3$.
As emphasized above, the freeze-out only applies to the heavy elements, while the non-equilibrium evolution of the light elements continues, in particular burning processes that can approach the solar distribution.
 During the decompression of hot and dense matter, fusion processes (burning) are generally possible after HEFO, where the nuclear statistical equilibrium of heavy nuclei is no longer maintained. This increases the abundance of light elements, and the freeze-out of light elements occurs at a later stage. In standard calculations, reaction networks are applied at this later stage. For example, in SN simulations, a nuclear statistical equilibrium is assumed as long as the temperature of a mass element is above 0.7 MeV, see, e.g., \citet{2017PASA...34...67F,ricigliano2024}.

\subsection{The level density of excited nuclei}

In the framework of our approach, where we have clustered matter at high temperatures, the intrinsic partition function of the nuclei $\{A,Z\}$ is an important ingredient.
A simple version was considered in \citet{Roepke:2024fzj}.
For the partial density $n_{AZ}(T,\mu_n,\mu_p)$ of the nucleus in the channel $\{A,Z\}$ the expression
\begin{eqnarray}
    \label{nAZ}
    n_{AZ}(T,\mu_n,\mu_p)&=&
    R_{AZ}(T,\mu_n,\mu_p)\, \left( \frac{2 \pi \hbar^2}{AmT}\right)^{-3/2}\nonumber\\
    &&\times \exp\left\{-\frac{E^0_{AZ}(T,\mu_n,\mu_p)-(A-Z)\, \mu_n-Z \mu_p}{T}\right\}
\end{eqnarray}
was given,
where $E^0_{AZ}(T,\mu_n,\mu_p)$ is the medium-modified ground state energy of the nucleus $\{A,Z\}$. 
The degeneracy factor $g_{AZ}$ and
the sum over all excited states, including the continuum contributions, are absorbed in the prefactor $R_{AZ}(T,\mu_n,\mu_p)$, the intrinsic partition function. For the light elements, the excited states of the nuclei and their degeneracy are known \citep{nuclei} 
so that the summation can be performed within the intrinsic partition function and the continuum contribution to the virial form  \citep{Ropke:2020hbm,Natowitz:2022npi}. For the heavier nuclei,
the summation over their excited states can be replaced by the integral over the density of states \citep{BohrM} 
\begin{eqnarray}
\label{RAZ}
    R_{AZ}=g_{AZ} +\frac{\pi^{1/2}}{12} \left(\frac{15\,\mathrm{MeV}}{A}\right)^{1/4}\int\limits_{E_\mathrm{min}}^{E_\mathrm{max}} dE\, E^{-5/4} \exp\left\{2\sqrt{\frac{E\, A}{15\,\mathrm{MeV}}}-\frac{E}{T}\right\},
\end{eqnarray}
where we take $E_\mathrm{min}\,{=}\,25\,\mathrm{MeV}/A$ and $E_\mathrm{max} $ as the binding energy of the bound state $\{A,Z\}$. 

A general expression for the nuclear level density $\rho(E)$ reads \citep{BohrM}
\begin{equation}
    \rho(E)=K_{\rm rot}K_{\rm vib}\rho_{\rm int}(E),
\end{equation}
where $K_{\rm rot}, K_{\rm vib}$ are the coefficients for rotational and vibrational enhancement of the non-collective internal nuclear excitations $\rho_{\rm int}(E)$. In this work, we take $K_{\rm rot}= K_{\rm vib}=1$.
Microscopic calculations of $\rho_{\rm int}(E)$ are rather complicated. 
Instead, usually an empirical approach is used, such as the back-shifted Fermi-gas model
\begin{equation}
    \rho_{\rm int}(E)= \frac{\pi^{1/2}}{12}a^{-1/4} (E-\Delta)^{-5/4} \exp\left[2 \sqrt{a(E-\Delta)}\right],
\end{equation}
where the backshift parameter is taken as the pairing energy $\Delta=12\,\chi /\sqrt{A}$ MeV, with $\chi=0, 1, -1$ for odd, even-even, odd-odd nuclei, respectively. 
The empirical parameter $a$ is approximated \citep{2024PhyS...99a5302S}
\begin{equation}
    \tilde a = \alpha A + \beta A^{2/3} = [ 0.073\,A + 0.195\,A^{2/3}]\,\, {\rm MeV}^{-1} .
\end{equation}
This liquid-drop result approximates the value $a=A/ 15$ MeV$^{-1}$ given in Eq. (\ref{RAZ}).
For a more detailed discussion of the intrinsic partition function see \citet{1997PhRvC..56.1613R} and \citet{1992NuPhA.543..517I}, see also \citet{Rauscher:2003ti} and, more recently, \citet{2021PhRvC.104d4319M}, \citet{2025NuPhA105823034O}.
The backshift $\Delta$ is not of relevance because we average over neighbored mass numbers $A$ so that the even-odd staggering are averaged out.

Another deviation from the liquid droplet model is the occurrence of magic numbers which is related to the shell structure of the single quasiparticle states in the mean-field nuclear potential. 
This additional contribution $\delta W(Z,N)$ to the binding energy has been parametrized in a semiempirical approach by \citet{1995PhRvC..52...23D}.
In \citet{Roepke:2024fzj} this correction was considered as rigid shift of the nuclear level density.
A more detailed description considers how the shell corrections change with excitation energy.
An empirical relation was considered by \citet{1992NuPhA.543..517I}. 
However, the freeze-out Lagrange parameters are far away from the states of matter which are available in recent experiments.

In this work we use the energy-dependent shell correction proposed by \citet{1992NuPhA.543..517I},  \citet{1997PhRvC..56.1613R},
\begin{equation}
\label{shellcorr}
   a(E,N,Z)= \tilde a \left[1+\delta W(Z,N)\frac{1-\exp[-\gamma (E-\Delta)]}{E-\Delta}\right],
\end{equation}
with $\gamma =0.05$ MeV$^{-1}$. This empirical value was obtained from a fit to known experimental data which refer to very different situations (lower energies, nearly symmetric matter) compared with the hot and neutron-rich state of matter considered here.  
Thus, the form of the level density of excited nuclei remains open. 
In addition, for a more adequate description, a nonequilibrium approach is required. However, this is beyond the scope of the present work.

\section{Phenomenological HEFO Lagrange parameters for stars}
\label{universality}

We now consider other stellar objects and  compare their composition with the solar abundance distribution of heavy elements.
An interesting phenomenon 
is the robust universality of the main $r$-process pattern, 
as pointed out in the Introduction and Sect. \ref{sec:universality_obs}.
Within our framework, almost the same Lagrange parameters can be used to characterize the HEFO conditions for the respective objects. 
A fit to the Lagrange parameters for the solar heavy element abundances is given in Tab. \ref{tab1}.
Compared with the calculations in \citet{Roepke:2024fzj},
we used the temperature dependence of the shell correction shifts, Eq. (\ref{shellcorr}), so that the Lagrange parameters for the solar distribution  are only marginally shifted.
\begin{table}[htp]
\begin{center}
\begin{tabular}{|c|c|c|c|c|c|}
\hline
&$\lambda_T$ [MeV] & $\lambda_n$ [MeV] & $\lambda_p$ [MeV] & $\log (\hat X_{96}) $ & $\log (\hat X_{356})$ \\
\hline
solar & 5.2904 & 940.294 & 845.055 & -6.826 & -8.0566 \\
$A$ & 5.30868 & 940.277 & 845.07 & -6.826 & -8.1566 \\
$B$ & 5.24664 & 940.336 & 844.935 & -6.926 & -8.1566 \\
$C$ & 5.20376 & 940.376 & 844.817 & -7.026 & -8.2566 \\
\hline
\end{tabular}
\end{center}
\caption{Lagrange parameters and variation of the abundances.}
\label{tab1}
\end{table}%

Universality is not a strict property of stars, but there exist deviations from a constant ratio of heavy-element abundances, valid for arbitrary $Z$. 
As discussed in Sect. \ref{sec:universality_obs}, an actinide-boost 
is observed for some stars, with an over-abundance of the heaviest elements.
On the other hand, stars are observed with an under-abundance of the rare-earth domain, if compared with the solar system abundances.
A variation of the abundances of elements can be described by a variation of the Lagrange parameters at HEFO.
In this section, our aim is to infer the Lagrange parameters  giving rise to the abundance patterns observed for several individual stars.\\

\begin{figure}[htbp]
\begin{center}
\includegraphics[width=0.6\linewidth]{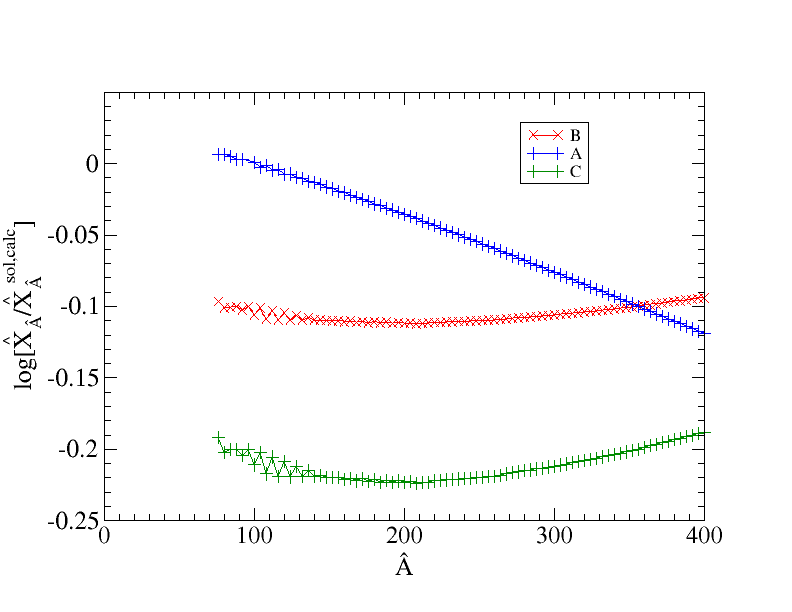}
\caption{Three different coarse-grained distributions of the heavy nuclei, $\hat A \ge 76$. The change $\log (\hat X_{\hat A})-\log (\hat X^{\odot,{\rm calc}}_{\hat A})$ relatively to the solar distribution is shown, for the parameter values of examples A, B, C, see Tab. \ref{tab1}. Temperature dependence of the shell corrections according Eq. (\ref{shellcorr}).} 
\label{test1}
\end{center}
\end{figure}

Before that, we study the effect of a change of the Lagrange parameters on the form of the distribution function.
We assume a change of Lagrange parameters to describe the modification of the heavy element distribution.  
With the normalization $\log[M_0]=0$ at given baryon density \citep{Roepke:2024fzj}, only two Lagrange parameters are free. 
We infer them from given values of $\hat X_{96}, \hat X_{356}$.
The corresponding Lagrange parameters are shown in Tab. \ref{tab1}.
Small changes of the Lagrange parameters give already substantial changes of the distribution, see Fig. \ref{test1}.

We see that the heavy element distribution is shifted downwards if the temperature $\lambda_T$ decreases, but the universality is approximately fulfilled if a minor change of the other parameters $\mu_n, \mu_p$ is performed.
We can also change the slope of the heavy element distribution function. If the slope of the distribution function is changed, universality is no longer valid.

For the comparison with individual stellar abundances, we have to use $Z$ instead of $A$ since the observed line spectra are attributed to the chemical elements. 
{In order to relate $Z$ to a mass number $A$, we take the line of stability in the neutron-proton plane.}
Note that we are considering double ratios so that different magnitudes of intervals attributed to a value of $Z$ will cancel out. In this work, we discuss various examples where observations of as many  elemental abundances as possible are available. 
{Our approach is phenomenological; we do not intend to describe various processes by which heavy elements can be formed, see the discussion in Sect. \ref{sec:3}.}

Of particular interest are low-metallicity stars observed in the halo of the Milky Way.
It is assumed that these are old stars, characterized by a low value of [Fe/H]. 
We use \citet{2012A&A...545A..31H}, Tabs. C1, C2, from which we only consider stars where data for all indicated elements are given, and perform the averages of [X/H] which gives the deviation from solar distribution. 
Values are shown in Tab.~\ref{Tab:stellarAb}.
Low metallicity is characterized by the [Fe/H] values. Dwarfs have masses of about 0.8 $M_\odot$, for the giant stars the mass 1 $M_\odot$ is assumed \citep{2012A&A...545A..31H}.
The error bars of the values of [X/H] are about 0.2.
There is a systematic shift of the ratios [X/H], increasing with $Z$ respective $A$.

\begin{figure}[htbp]
\begin{center}
\includegraphics[width=0.6\linewidth]{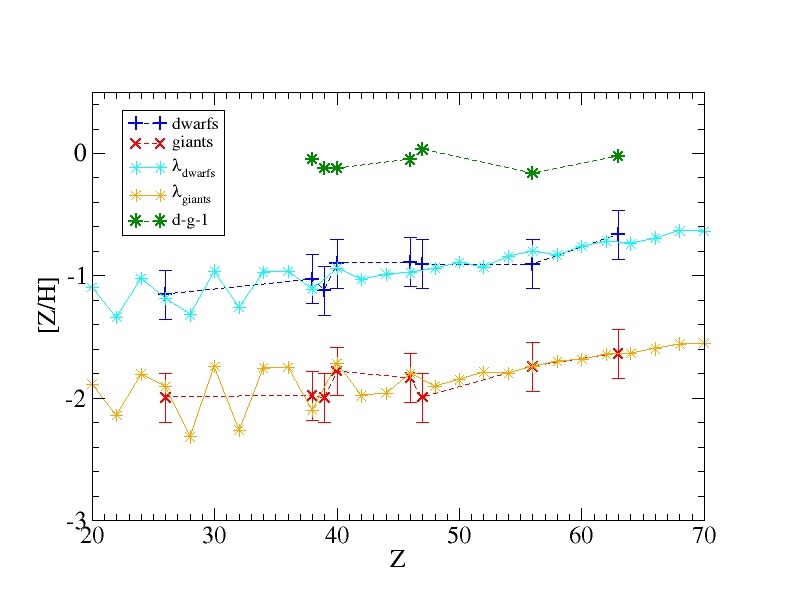}
\caption{Stellar abundances [X/H] for halo dwarfs and giants  according \citet{2012A&A...545A..31H}. The Lagrange parameters $\lambda_{\rm dwarfs}$ and $\lambda_{\rm giants}$ are given in Tab. \ref{Tab:Lagrange}. In addition, the difference d-g-1 between both averages (dwarfs-giants) is shown, after subtraction of 1 for convenience.} 
\label{Fig:stellarAb}
\end{center}
\end{figure}

This can be reproduced using appropriate Lagrange parameters (we use $T,\mu_n,\mu_p$ instead of $\lambda_T,\lambda_n,\lambda_p$). 
A fit is shown in Fig.~\ref{Fig:stellarAb} and Tab.~\ref{Tab:Lagrange}. 
The lower metallicities correspond to lower $T$ values. 
In Fig.~\ref{Fig:stellarAb}, the light elements have larger scatter but are not relevant here since 
we consider only the heavy elements ($Z >30$).

\begin{table}[htp]
\begin{center}
\begin{tabular}{|c|c|c|c|c|c|c|c|c|c|c|}
\hline
No.& mass & [Fe/H]& [Sr/H]& [Y/H]& [Zr/H]& [Pd/H]& [Ag/H]& [Ba/H]& [Eu/H] \\
\hline
1 & dwarfs  & -1.153 & -1.028 & -1.122 & -0.902 & -0.889 & -0.962 & -0.904 & -0.662 \\
2 & giants  & -1.995 & -1.979 & -1.998 & -1.781 & -1.836 & -1.994 & -1.742 & -1.638 \\
\hline
\end{tabular}
\end{center}
\caption{Stellar abundances according to \citet{2012A&A...545A..31H}, averages for dwarfs and giants.}
\label{Tab:stellarAb}
\end{table}%

We obtain different Lagrange parameters for dwarfs and giants.
The metallicity for giants is smaller, the Lagrange parameter $T$ is smaller.
However, the universality of the heavy elements is fulfilled, the difference between both groups shown in Fig. \ref{Fig:stellarAb} is nearly independent on $Z$.

Nevertheless, both curves for [X/H], dwarfs and giants, as function of $Z$ show a slope which is in conflict with the universality with the solar distribution, i.e., they are not constant, independent on $Z$. 
Such a slope can be described in our phenomenological approach by fitted values of the Lagrange parameters.

\begin{table}[htp]
\begin{center}
\begin{tabular}{|c|c|c|c|}
\hline
&$T$ [MeV] & $\mu_n$ [MeV] & $\mu_p$ [MeV]  \\
\hline
solar & 5.2904 & 940.294 & 845.055  \\
dwarfs & 4.6391 & 940.875 & 843.873  \\
giants &4.3576& 941.101 & 843.166 \\
Honda et al.&4.5551 & 940.944 & 842.349  \\
\hline
\end{tabular}
\end{center}
\caption{Lagrange parameters, dwarfs and giants according \citet{2012A&A...545A..31H},  are fitted to the data of Tab. \ref{Tab:stellarAb}, see Fig. \ref{test1}. Lagrange parameters are also given for two stars with drop-offs through the rare-earth domain according \citet{2007ApJ...666.1189H}.}
\label{Tab:Lagrange}
\end{table}%

\begin{figure}[htbp]
\begin{center}
\includegraphics[width=0.6\linewidth]{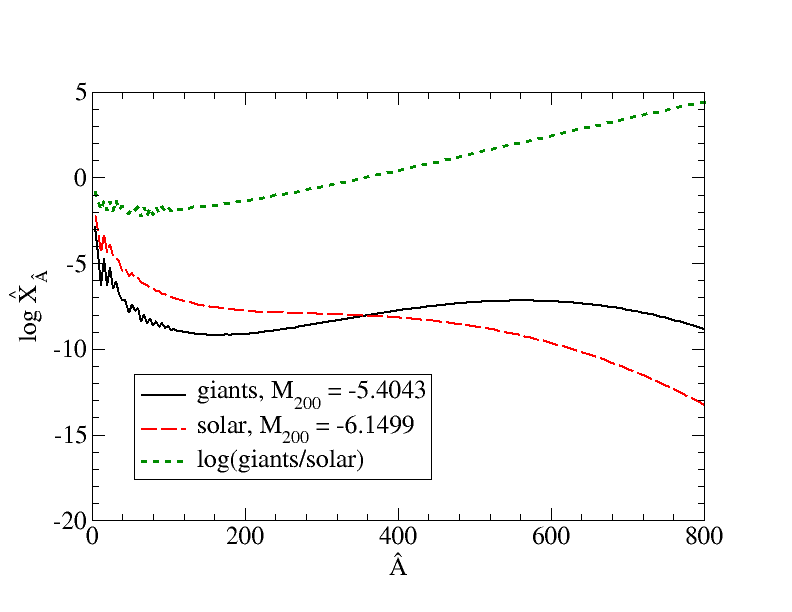}
\caption{HEFO primordial accumulated mass fraction distribution $\log \hat X_{\hat A}$. The Lagrange parameters $\lambda_{i}^\odot$ and $\lambda^{\rm giants}_i$ are given in Tab. \ref{Tab:Lagrange}. 
In addition, the ratio of both (giants/solar) is also shown.} 
\label{Fig:giants}
\end{center}
\end{figure}

For the stars shown in Fig.~\ref{Fig:stellarAb}, the ratio of heavy elements increases with $Z$ in the metal-poor giants compared with the solar system. To discuss this point, we revert to the distribution with respect to the mass number $A$.
To show the general behavior of the primordial accumulated mass fraction distribution $\hat X_{\hat A}$, we perform the calculation up to $A=800$, see Fig.~\ref{Fig:giants}. 
Whereas the region $44 \le \hat A \le 172$ corresponds to the results given in Fig.~\ref{Fig:stellarAb}, we observe an interesting trend at large values of $\hat A$.
The mass fraction is higher than the solar value. These superheavy nuclei will decay (fission, $\alpha$-decay) and feed the heaviest stable elements.
We predict that these stars, described in \citet{2012A&A...545A..31H}, will have an enhanced abundance of the heaviest elements (3rd peak) since the decay products of the superheavy-element primordial distribution at HEFO will populate these elements.
In Fig. \ref{Fig:giants} we give the total amount of material $M_{200}$ found in primordial nuclei with $A \ge 200$.
The value for the solar Lagrange parameters it is $\log\left[M_{200}^\odot\right]= - 6.1499$.
For the parameter values $\lambda_i$ of giants, given in Tab.~\ref{Tab:Lagrange}, $\log\left[M_{200}^{\rm giants}\right]=-5.4043$, about 5 times larger than solar.

An actinide-boost halo star (metal-poor giant star HE 2252-4225, [Fe/H]$\,{=}\, -2.63$) was studied by \citet{2014A&A...569A..43M}. 
 In their Fig.~3, the overabundance of the heaviest elements ($60 \le Z \le 90$) is clearly seen, possibly originating from radioactive decay of superheavy elements. 
However, \citet{2015EPJA...51...22G} point out that fission and its consequences for nucleosynthesis observables remain an open problem due to a lack of empirical data.

As mentioned in Sect. \ref{sec:universality_obs}, the chemical analysis of
52 stars was performed with high resolution by \citet{2025A&A...693A.294A}. An extreme overabundance of the elements of the third peak
was observed in the Eu-poor stars, which, according to \citet{2025A&A...693A.294A} supports the picture of a variable $r$-process as opposed to a universal outcome. They interpret this finding as an indication for an additional early-time $r$-process contribution and point out that none of the currently considered astrophysical sites can fully explain the increased abundances of Os, Ir, and Pt.
Therefore, it is of fundamental interest to investigate the physical conditions an astrophysical site has to reach in order to explain such a new $r$-process. 

Likewise, as also mentioned in Sect.~\ref{sec:universality_obs}, \citet{2021RvMP...93a5002C} point out that increased
abundances of Th and U compared to lighter $r$-process elements are also found in other stars.
The authors take the observed actinide enhancement in some stars
with metallicities [Fe/H] ${\approx}\, -3$ as an indication for a non-universal $r$-process, which played a role in very early galactic evolution proceeded under varying conditions depending on the $r$-process site. 
Our aim is to express these varying conditions by varying the parameter values of the Lagrange parameters $\lambda_i$, as indicated in Fig. \ref{Fig:giants}.

\begin{figure}[htbp]
\begin{center}
\includegraphics[width=0.6\linewidth]{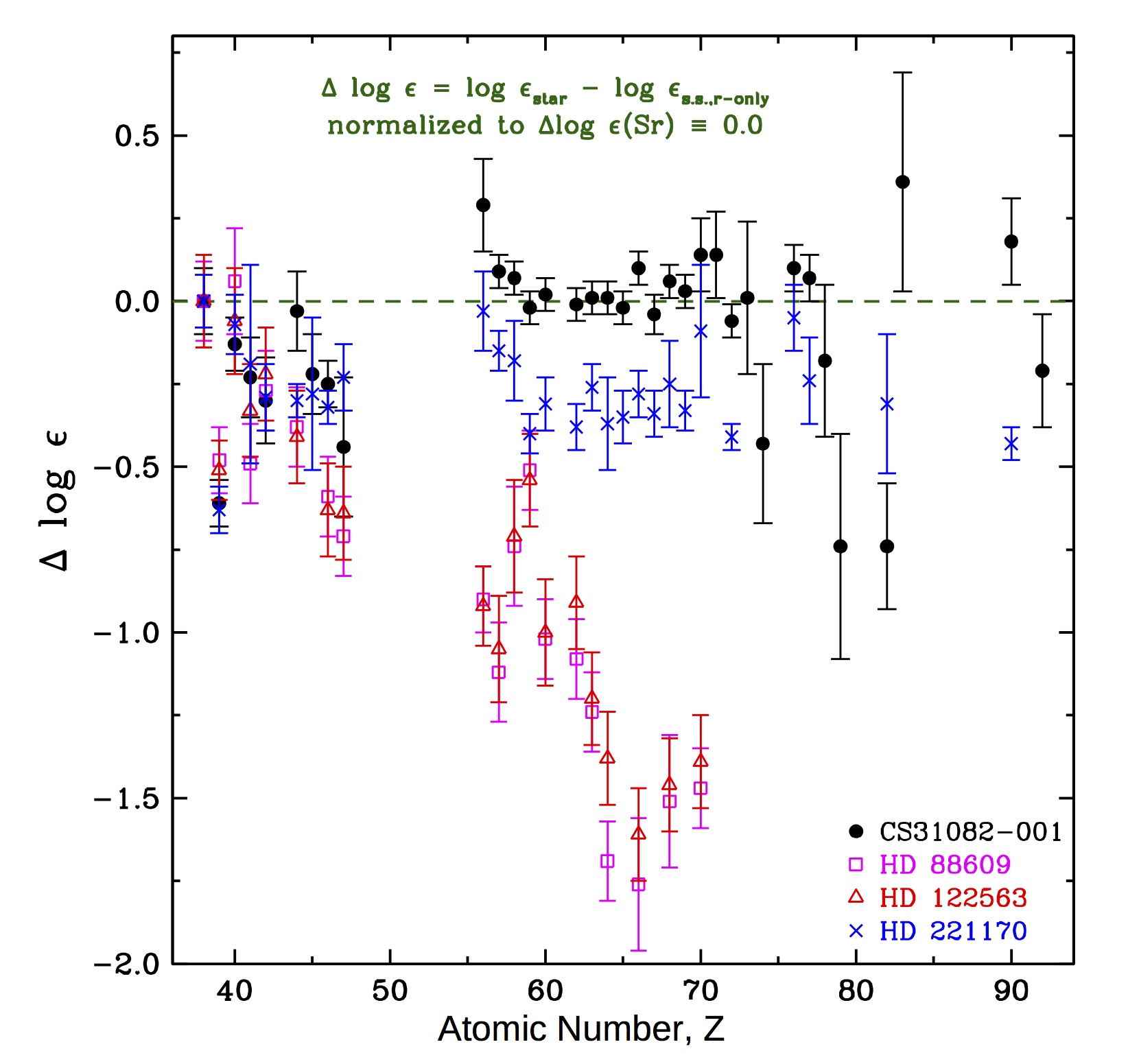}
\caption{Fig. 4 from \citet{2021RvMP...93a5002C} with permission from the Publisher. For details see the text.}
\label{honda}
\end{center}
\end{figure}

\begin{figure}[htbp]
\begin{center}
\includegraphics[width=0.6\linewidth]{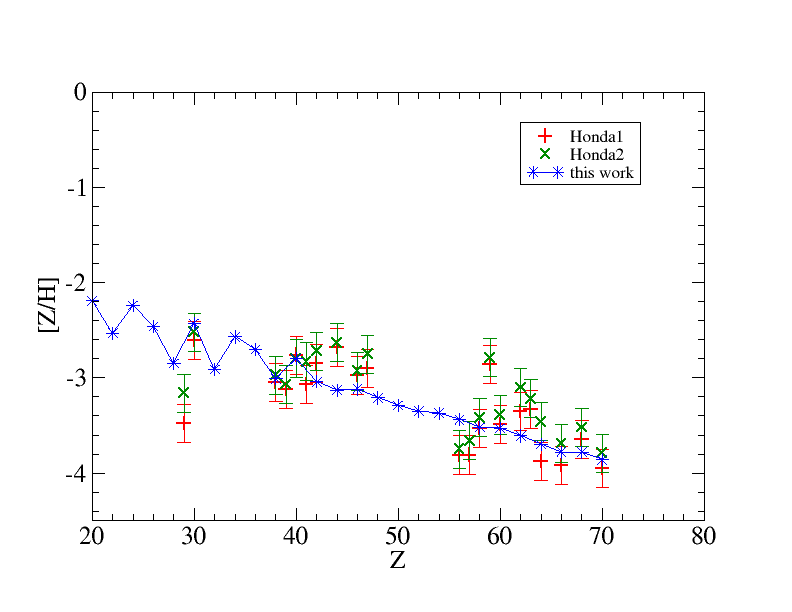}
\caption{Abundance pattern for two stars (+ HD 88609, x HD122563) of \citet{2007ApJ...666.1189H} compared to results from fit of Lagrange parameters given in Tab. \ref{Tab:Lagrange} (blue asterisk connected by line to guide the eyes).} 
\label{hondaLagr}
\end{center}
\end{figure}

For several low-metallicity stars, a reverse strong deviation from universality was observed.
We consider the drop-offs across the rare-earth domain mentioned in Sect. \ref{sec:universality_obs}, 
see Fig. \ref{honda}, taken from \citealt{2021RvMP...93a5002C}.
The stellar abundance sets are CS31082-001 \citep{2013A&A...550A.122S}, HD 88609, HD 122563 \citep{2007ApJ...666.1189H}, and HD
221170 \citep{2006ApJ...645..613I}.
From $Z=55$ to $70$, a significant decrease in abundance can be observed for some stars.
It is argued that these stars do not share the same chemical enrichment history as the others in the ensemble, but originate from a different source \citep{2014ApJ...797..123H}.
We have determined the Lagrange parameters for these stars, see Table \ref{Tab:Lagrange}.
The abundances calculated from the Lagrange parameters are shown in Fig. \ref{hondaLagr} in comparison to the observed distribution.
We can reproduce the general trend of the distribution of heavy elements with the Lagrange parameters from Table \ref{Tab:Lagrange}.
However, as can be seen in Fig. \ref{honda}, scatter and error bars are large.

\begin{table}[htp]
\begin{center}
\begin{tabular}{|c|c|c|}
\hline
$Z$  & [X/Fe] & [X/Fe]  \\
& HD 88609 & HD 122563 \\
\hline
[Fe/H]  &  -3.0  &  -2.7 \\
\hline
29& -0.48&  -0.46\\
30& 0.39&    0.18\\
38& -0.05&  -0.27\\
39& -0.12&  -0.37\\
40& 0.24&   -0.10\\
41& -0.07&  -0.13\\
42& 0.15&   -0.02\\
44& 0.32&    0.07\\
46& 0.03&   -0.23\\
47& 0.10&   -0.05\\
56& -0.81&  -1.05\\
57& -0.81&  -0.96\\
58& -0.53&  -0.72\\
59& 0.14&   -0.09\\
60& -0.49&  -0.69\\
62& -0.35&  -0.40\\
63& -0.33&  -0.52\\
64& -0.88&  -0.76\\
66& -0.92&  -0.99\\
68& -0.65&  -0.82\\
70& -0.95&  -1.09\\
\hline
\end{tabular}
\end{center}
\caption{Two stars with drop-offs through the rare-earth domain described by \citet{2007ApJ...666.1189H}.}
\label{Tab:Honda}
\end{table}%

Instead of a strict universality of the distribution of heavy elements, we find variability of the distribution. 
This leads to the variability of the three Lagrange parameters $\lambda_T, \lambda_n, \lambda_p$, which represent the non-equilibrium generalizations of $T,\mu_n,\mu_p$. 
We consider a phenomenological approach to show that the distribution of heavy elements in different stars is related to these Lagrange parameters. 
In this section, we do not provide an interpretation of the astrophysical site where these parameters occur. 
The existence of these Lagrange parameters is valuable in itself.
They characterize every astrophysical object, especially stars, in terms of the distribution of heavy elements.

\section{Discussion}
\label{sec:3}
Within the framework of the non-equilibrium freeze-out approach, we obtain values for the Lagrange parameters, which are the non-equilibrium generalisations of $T, \mu_n, \mu_p$. 
{Our conditions \citep{Roepke:2024fzj} ($T\approx 5\, {\rm MeV}\, \cong 5.8 \times 10^{10}$ K,  
$n_B\approx 0.013$ fm$^{-3} \cong 2.2 \times 10^ {13}$ g/cm$^3$, $Y_p\approx 0.13$,} see Table \ref{Tab:Lagrange}) are found in supernova explosions,
in binary NS mergers and in similar astrophysical objects with high energy density. 
{For example, \citet{2014EPJA...50...46F,2017PASA...34...67F} and  \citet{2023A&A...672A.160D}
provided examples where these parameters for density and temperature occur in connection 
with supernova simulations and in the crust of proto-neutron stars.
Undoubtedly, such conditions can occur in special astrophysical sites,
and the production of heavy $r$-process elements has been detected in connection with the NS merger event GW170817 \citep{watson2019a}.
There are numerous studies,
 see, e.g., \citet{2014AIPC.1594..266W}, \citet{2015MNRAS.448..541J}, \citet{2019Natur.569..241S}, and \citet{2017ARNPS..67..253T}, which show that $r$-process elements can be produced from compact object mergers.
In these publications, the calculation of nucleosynthesis begins when the temperature falls below 1 MeV, which is significantly below HEFO temperatures. 
The proton fractions $Y_p$ with a value of about 0.1–0.15 are in the same range as for HEFO.
Since we are not concerned with the astrophysical origin of heavy elements in this article, we will not discuss where the calculation of nucleosynthesis should begin, at HEFO temperatures or at temperatures of about 0.5 MeV.
While HEFO  has an initial distribution of heavy elements from the outset, conventional standard approaches must generate the heavy elements using nuclear reaction networks.}

Postprocessing the hydrodynamical evolution, nuclear reaction networks such as SkyNet \citep{Lippuner:2017tyn} or WinNet \citep{reichert2023a} are usually applied to describe the evolution of the chemical composition, assuming NSE for the distribution of elements also for temperatures $T\sim 0.5$ MeV as initial condition. Under certain conditions, $r$-abundance pattern are obtained, see \citet{Wanajo:2007jj,2015MNRAS.448..541J,deJesusMendoza-Temis:2014owk,2023ApJ...951L..12J,2025A&ARv..33....2B}.
However, the uniformity of the heavy element distribution is not easily explained, since the result of reaction kinetics depends on the duration and other characteristics of the neutron flux which is required to run the $r$ process.
In contrast,  HEFO is able to reproduce the universality feature as a consequence of decompression from a hot and dense state of matter described by only a few Lagrange parameters.

{Recent studies have shown that actinide-boosted star signatures can also be reproduced in specific astrophysical contexts, 
for example, by the intermediate neutron capture process ($i$-process) that occurs during proton-ingestion events (PIEs) in AGB stars with low metallicity (e.g., \citet{2022A&A...667L..13C,2024Galax..12...66C,2025EPJA...61...68C,2010JPhCS.202a2024K} ).
Recent work by \citet{2024ApJ...966...11P} shows that neutrino-driven outflows in core-collapse supernovae can reproduce the observed overproduction of lighter $r$-process elements between Sr and Ag in very metal-poor stars such as HD 122563 through a combination of weak $r$-process and $\nu p$-process nucleosynthesis. However, we do not intend to discuss different pathways for the formation of these elements \citep{2023MNRAS.523.2126P,2026enap....2..744K}, but merely wish to provide a phenomenological description of the distribution.
}

It is an open question whether events such as supernova explosions or binary NS mergers are the only sites for the origin of the heavy elements in our Universe, see  \citet{2018ApJ...855...99C}, \citet{2021MNRAS.505.5862W}, and further articles cited in the Introduction and in Sect. \ref{sec:universality_obs}. 
A major problem is the appearance of heavy elements already in the low-metallicity stars which are assumed to be very old. 
In particular, a large ratio [Eu/Fe] has been observed in stars with [Fe/H]$<-2$, see \citet{Wehmeyer:2015sra,Wehmeyer:2019ovu,2020ApJ...898...40C,2024ApJ...971..158R}, and \citet{2025ApJ...985..154C}. 

It is assumed that metallicity can be used as an indicator of the age of a star. Low-metallicity stars ([Fe/H] $\le$ -2.5) represent early stars, so metallicity serves as a clock in the context of galactic chemical evolution (GCE) to determine the time at which stars formed from galactic matter. 
An overview of the research field of GCE would go beyond the scope of this article, so we refer to \citet{Matteucci2012}. 
Here we only mention the following problem:
Stars with high $r$-process abundances ($0\, {<} \,\mathrm{[Eu/Fe]}\, {<} \, 2$) and extremely low metallicity  ($\mathrm{[Fe/H]}\,{<}\, -2.5$) have been observed, and an astrophysical site for nucleosynthesis is under discussion in an environment with lower metallicity than binary NS mergers could have, see \citet{Wehmeyer:2015sra,Wehmeyer:2019ovu,2015A&A...577A.139C,2019MNRAS.483.5123H,2022EPJWC.26009002T,2022A&A...663A..70F}.
{Models of GCE were formulated trying to explain the early $r$-process onset, see, e.g., \citet{2015ApJ...814...41H}, \citet{2015ApJ...807..115S}, \citet{2015MNRAS.447..140V,2015ApJ...804L..35I}, \citet{2018ApJ...865...87O}, \citet{2018IJMPD..2742005H}, and \citet{2019Natur.569..241S}. 
The issues related to GCE will be discussed in forthcoming work.}

While the occurrence of stars with low metallicity and high content of $r$-elements is difficult to understand using standard approaches, it is consistent with the freeze-out scenario with Lagrange parameters obtained from the solar distribution, see Fig. \ref{fig:XA1}. A ratio [Eu/Fe]$\,{=}\,3$ is obtained at the corresponding freeze-out conditions, with [Fe/H]$\,{=}\,-3$.
Since in the HEFO scenario only the heavy element distribution freezes out, and the excited heavy nuclei decay after freeze-out,
the light elements evolve further after HEFO, for instance due to burning processes. 
Therefore, with increasing amount of Fe, smaller values for the ratio [Eu/Fe] appear. 
Since we are not concerned in this work with the nuclear reaction processes after HEFO for the light elements such as Fe, we only mention the distribution of the oldest stars with a ratio [Eu/Fe]$ \le 2.45$, see \citet{2020ApJ...898...40C,2024ApJ...971..158R}. 

{Although the production of $r$-process elements in NS mergers has been confirmed, it remains unclear whether they merely contribute to the enrichment of the Universe with heavy elements or whether they can account for the total cosmic abundance of heavy nuclei. 
An interesting possibility is the inhomogeneous Big Bang nucleosynthesis (IBBN), see \citep{Gonin:2025uvc} that is based on the 
concept of HEFO \citep{1987PhLB..185..281R,Roepke:2024fzj}.
Even before Big Bang nucleosynthesis, primordial black holes (PBH) and other large-scale density fluctuations are assumed to exist. Such density fluctuations, which survive the homogeneous Big Bang nucleosynthesis (HBBN) time scale, are possible sites where heavy element nucleosynthesis can take place.
In constrast to scenarios that postulate events such as magnetorotational SNe, collapsars/hypernovae and possibly binary compact object mergers to take place extremely early, hot and dense matter is present in the IBBN from the beginning and must not created by accretion from low-density, metal-less population III matter as in the HBBN.
We do not provide detailed calculations of the distribution of inhomogeneities and their lifetime here.
The discussion of a cosmological scenario that includes PBHs and other density fluctuations would go beyond the scope of the present work.}

{Values can be specified for the Lagrange parameters of HEFO that describe the frequently discussed large [Eu/Fe] ratio in stars with low metallicity. 
Assuming that low [Fe/H] values indicate that these stars formed early, the heavy elements should also have been formed in an early process. 
For stars with [Eu/Fe] $> 0.3$, a 
behavior is shown that points to the appearance of a further production site, see, e.g., \citet{2023ApJ...958...45K} and \citet{2025arXiv250318233F}.
An open question is whether there are several kinds of sources.
Neutron-star mergers as the source of $r$-process-enhanced metal-poor stars in the Milky Way are considered by \cite{2019ApJ...872..105S, 2019ApJ...876...28S}.  
They show that even when we adopt the $r$-process yield estimates
observed in GW170817, neutron-star mergers by themselves can only explain the observed frequency of $r$-process-enhanced
stars, if the birth rate of DNSs per unit mass of stars is boosted to $\approx 10^{-4} M_\odot^{-1}$.
The investigation of \citet{2019MNRAS.483.5123H} lead to the conclusion that 
neither electron capture supernovae or neutrino-driven winds are able to adequately explain the
observed Eu levels 
(for a general discussion of galactic evolution, see also \cite{2025arXiv250620436K}).
Large scatter of data is also shown in  \citet{2024ApJ...971..158R}.
There must be another source for the heavy element production in the early Universe, as pointed out, e.g., by \citet{Wehmeyer:2015sra,Wehmeyer:2019ovu,2014A&A...565A..51C}, \citet{2024ApJ...971..158R}, \citet{2025ApJ...985..154C}, and \citet{2025RAA....25h5015A}. \citet{2025A&A...693A.294A} emphasize the challenge their results present to conventional nucleosythesis scenarios and the need for an additional early production channel for $r$-process elements that does not require mergers of compact objects \citep[see also][]{2025arXiv250806020S}.
This view is also supported
by the GCE models of  \cite{Cote:2018qku}, that suggest an extra $r$-process site to provide $r$-process enrichment in the early
Universe.} 

{The possibility of an early, previously unknown process of nucleosynthesis is the subject of intense debate in the literature. Our work does not attempt to answer the question of the astrophysical scenario for such a process. 
We merely point to one possibility, namely the existence of very early, primordial fluctuations in density and temperature. 
The values for the Lagrange parameters and their dispersion presented here may provide an indication of the properties of such an early process of nucleosynthesis.
}

\section{Conclusions}
\label{conclusion}

{In the Heavy Element Freeze-Out (HEFO) model, Lagrange parameters $\lambda_i$ are introduced to characterize the distribution of heavy elements. Specific parameter values can be determined from the distribution of heavy elements in different stars.
Deviations from the uniformity of the $r$-process element abundance can be mapped to a scatter in the values of the Lagrange parameters that characterize the conditions under which the HEFO occurs.
Larger variability in the Lagrange parameters is observed in stars with low metallicity. We have presented several examples, such as actinide-boost stars and heavy-element drop-off distributions. 
A three-parameter HEFO framework can reproduce the coarse-grained $r$-process abundance patterns, and  modest parameter variations can fit the observed stellar diversity.}

\section*{Funding}
D.B.\ was supported by the Polish NCN
under grant No. 2021/43/P/ST2/03319. 
G.R.\ acknowledges a honorary stipend from the Foundation for Polish Science within the Alexander von Humboldt program under grant No. DPN/JJL/402-4773/2022. The work of F.K.R.\ is supported by the Klaus Tschira Foundation, by the Deutsche Forschungsgemeinschaft (DFG, German Research Foundation) -- RO 3676/7-1, project number 537700965,
and by the European Union (ERC, ExCEED, project number 101096243). Views and opinions expressed are, however, those of the authors only and do not necessarily reflect those of the European Union or the European Research Council Executive Agency. Neither the European Union nor the granting authority can be held responsible for them.

\section*{Acknowledgments}
We thank Benjamin Wehmeyer for his comments after careful reading of this manuscript and Tobias Fischer for his valuable discussions. 

\section*{Data Availability Statement}
The datasets generated for this study are available from the RODARE database under DOI 10.14278/rodare.4149.

\bibliographystyle{Frontiers-Harvard} 


\end{document}